\documentclass[aps,twocolumn,prb,preprintnumbers,amsmath,amssymb,superscriptaddress,floats]{revtex4-2}

\usepackage{graphicx}
\usepackage{bm}

\usepackage{natbib}
\usepackage{float}
\usepackage{multirow}
\usepackage{array}
\usepackage{tikz}
\usetikzlibrary{tikzmark}
\usepackage{physics}
\usepackage{siunitx}
\usepackage[colorlinks=true, urlcolor=blue, citecolor=black, linkcolor=black, pdfborder={0 0 0}]{hyperref}

\def\pzo{$\mathrm{Pr}_2\mathrm{Zr}_2\mathrm{O}_7$}
\def\pio{$\mathrm{Pr}_2\mathrm{Ir}_2\mathrm{O}_7$}
\def\cm{cm$^{-1}$}

\begin{document}

\title{Phonon spectrum of \pzo\ and \pio\ as an evidence of coupling of the lattice with electronic and magnetic degrees of freedom}

\author{Yuanyuan Xu}
\affiliation{Institute for Quantum Matter and Department of Physics and Astronomy, Johns Hopkins University, Baltimore, Maryland 21218, USA}

\author{Huiyuan Man}
\affiliation{Institute for Quantum Matter and Department of Physics and Astronomy, Johns Hopkins University, Baltimore, Maryland 21218, USA}
\affiliation{Institute for Solid State Physics, University of Tokyo, Kashiwa, Chiba 277-8581, Japan}

\author{Nan Tang}
\affiliation{Institute for Solid State Physics, University of Tokyo, Kashiwa, Chiba 277-8581, Japan}

\author{Takumi Ohtsuki}
\affiliation{Institute for Solid State Physics, University of Tokyo, Kashiwa, Chiba 277-8581, Japan}

\author{Santu Baidya}
\affiliation{Department of Physics and Astronomy, Rutgers University, Piscataway, New Jersey 08854-8019, USA}

\author{Satoru Nakatsuji}
\affiliation{Institute for Quantum Matter and Department of Physics and Astronomy, Johns Hopkins University, Baltimore, Maryland 21218, USA}
\affiliation{Institute for Solid State Physics, University of Tokyo, Kashiwa, Chiba 277-8581, Japan}
\affiliation{Department of Physics, University of Tokyo, Bunkyo-ku, Tokyo 113-0033, Japan}
\affiliation{CREST, Japan Science and Technology Agency, Kawaguchi, Saitama 332-0012, Japan}
\affiliation{Trans-scale Quantum Science Institute, University of Tokyo, Bunkyo-ku, Tokyo 113-0033, Japan}

\author{David Vanderbilt}
\affiliation{Department of Physics and Astronomy, Rutgers University, Piscataway, New Jersey 08854-8019, USA}

\author{Natalia Drichko}
\affiliation{Institute for Quantum Matter and Department of Physics and Astronomy, Johns Hopkins University, Baltimore, Maryland 21218, USA}
\email{Corresponding author. Email:drichko@jhu.edu}

\begin{abstract}

Magnetic materials with pyrochlore crystal structure form exotic magnetic states due to the high lattice frustration. In this work we follow the effects of coupling of the lattice and electronic and magnetic degrees of freedom in two praseodymium-based pyrochlores  \pzo\ and \pio. In either of these materials, the presence of magnetic interactions does not lead to magnetically ordered low temperature states, however their electronic properties are different. A comparison of Raman phonon spectra of \pzo\ and \pio\ allows us to identify magneto-elastic coupling in \pzo\ that elucidates its magnetic properties at intermediate temperatures, and allows us to characterize phonon-electron scattering in the semimetallic \pio. We also show that the effects of random disorder on the Raman phonon spectra is small.

\end{abstract}

\maketitle
\date{\today}%

\section{Introduction}

Frustrated magnetism and the search for spin liquid materials is one of the main topics of condensed matter physics in recent years. The pyrochlores crystal structure provides a geometrically frustrated lattice suitable for hosting classical and quantum spin-ice and spin liquid states~\cite{Gardner2010,Gingras2014,Chen2016}.

In this work we present a phonon Raman scattering study of praseodymium-based pyrochlores  \pzo\ and \pio. In both materials  Pr$^{3+}$ is in the magnetic $J = 4$ state, and in \pio\ the Ir$^{4+}$ is also magnetic with $J_{\mathrm{eff}} = 1/2$. Despite the presence of magnetic interactions, the highest quality samples do not order magnetically down to the lowest measured temperatures for either of these compounds. These two materials are rarely discussed together due to the drastically different electronic properties. \pzo\ is a band insulator and a quantum spin ice candidate~\cite{Kimura2013}. \pio\ shows metallic resistivity which flattens below 50~K  due to a Kondo effect~\cite{Nakatsuji2006}. This material is  suggested to be a Luttinger semimetal, that is to host a quadratic band touching at the $\Gamma$ point of the Brillouin zone (BZ)~\cite{Cheng2017}. Early on it was suggested to be a chiral metallic spin liquid~\cite{Nakatsuji2006,Flint2013,Ni2021,Yao2018}. We found that a comparative study of their phonon spectra can reveal information of the coupling of the lattice to magnetic and electronic degrees of freedom which is difficult to identify otherwise.

While these two compounds are relatively well studied, the origin of their low temperature states is still under discussion. In both of these materials, structural disorder was claimed to influence the low temperature state~\cite{Maclaughlin2015,Rau2019,Kimura2013,Martin2017}, but other subtle effects related to the magneto-elastic interactions and phonon-electron scattering for metallic \pio\ can  be important.

In this study we use  phonon Raman spectroscopy to study \pzo\ and \pio. Raman scattering has shown its high sensitivity to magneto-elastic coupling, small lattice distortions, and structural disorder\cite{Drichko2016}. Somewhat unexpectedly, we do not find any significant effects of random disorder on the Raman spectra of these materials. For both materials, the only phonons that show broadening and splitting are $E_g$ phonons, which can be associated with the tilting of octahedra as discussed in Ref.~[\onlinecite{Trump2018}]. A very similar phonon spectrum originated from the similar crystal structures but different electronic properties allow us to identify phonon-electron scattering in \pio, and effects of magneto-elastic coupling in \pzo.

\section{Experimental}

Polarized Raman scattering spectra were measured from (111) surface of the \pzo\ and \pio\ single crystals grown by the floating zone method and the flux method respectively~\cite{Koohpayeh2014,Millican2007}. Raman scattering was excited with 514.5~nm line of Ar$^{+}$-Kr$^{+}$ mixed gas laser. Raman spectra were measured using the Jobin-Yvon T64000 triple monochromator spectrometer with a liquid nitrogen cooled CCD detector in the temperature range from 300 down to 14~K, and in frequency range from 800 down to about 30~\cm.

 The pyrochlore lattice has $Fd\bar{3}m$ (No.~227) space group corresponding to $O_h$ point group. The polarization-resolved spectra were measured in $(x,x)$ and $(x,y)$ geometries, where $x$ denotes an arbitrary direction in the (111) plane and $y$ is perpendicular to $x$. In Tab.~\ref{table:intensity} we present what symmetries of scattering channels are observed in the two measured geometries.

\begin{table}[H]	
	\centering
	\caption{Components of  Raman tensor  for $(x,x)$ and $(x,y)$ polarizations.\label{table:intensity}}
	\begin{ruledtabular}
	\begin{tabular}{
		>{\raggedright\arraybackslash}p{0.2\linewidth}
		>{\centering\arraybackslash}p{0.2\linewidth}
		>{\centering\arraybackslash}p{0.2\linewidth}
		>{\centering\arraybackslash}p{0.2\linewidth}}

		Geometry & $A_{1g}$ & $E_{g}$ & $T_{2g}$\\
		\hline
		$(x,x)$ & $a^2$ & $b^2$ & $c^2$\\
		$(x,y)$ & 0 & $b^2$ & $\frac{2}{3}c^2$
	\end{tabular}
	\end{ruledtabular}
\end{table}

\begin{table}[H]
	\centering
	\caption{Wyckoff positions and $\Gamma$ point representations for \pzo\ (\pio). \label{table:wyckoff}}
	\begin{ruledtabular}
	\begin{tabular}{
		>{\raggedright\arraybackslash}p{0.2\linewidth}
		>{\centering\arraybackslash}p{0.3\linewidth}
		>{\centering\arraybackslash}p{0.4\linewidth}}
		Element & Wyckoff positon & $\Gamma$ representation \\
		\hline
		Pr & $16c$ & Inactive \\
		Zr (Ir) & $16d$ & Inactive \\
		O & $48f$ & $A_{1g} + E_g + 3T_{2g}$ \\
		O$'$ & $8a$ & $T_{2g}$ \\
	\end{tabular}
	\end{ruledtabular}
\end{table}

\section{Results}

\begin{figure}[h]
	\includegraphics[width=\linewidth]{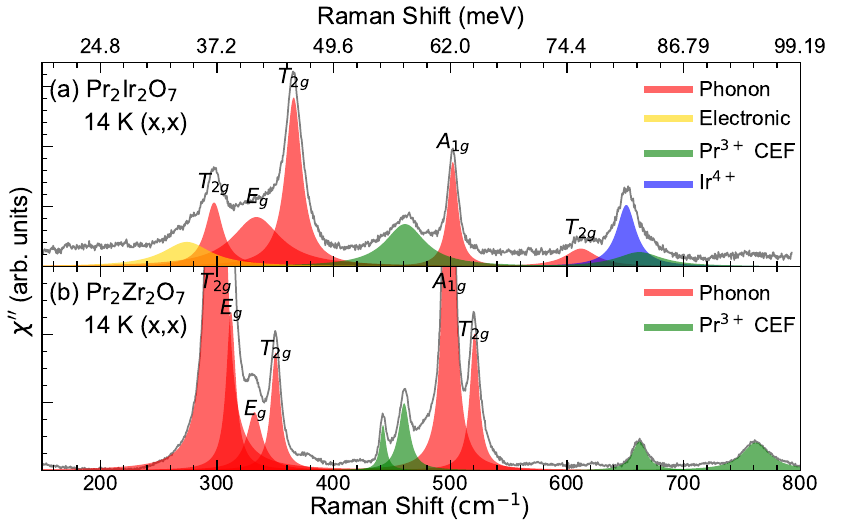}
	\caption{Raman scattering spectra of (a) \pio\ and (b) \pzo\ at 14 K in parallel polarization configuration. }
	\label{excitations}
\end{figure}

Raman scattering spectra of the single crystals of  \pzo\ and \pio\ in the spectral range between 150 and 800~\cm\ in $(x,x)$ scattering channel at $T = 14$~K are presented   in Fig.~\ref{excitations}. The low temperature data allows to resolve not only phonon excitations, but also crystal electric field (CEF) which can be observed only at low temperatures~\cite{Xu2021}. This spectral range covers all Raman-active phonons and excitations except the lowest excited state at about 80~\cm\ discussed in Ref.~\cite{Xu2021}. Due to the similarities of the crystal structure, many of the observed excitations show similar energies in these two compounds, however the Raman response of \pio\ is approximately two orders of magnitude weaker than that of \pzo, due to the fact that the first material is metallic. The screening present in the metallic state also affects the relative intensity of the phonons and  CEF excitations. For the latter, in \pio\ spectra we observe only the CEF exitation at around 460 \cm, which is coupled to phonons by a vibronic coupling process\cite{Xu2021}. Raman-active phonons  (marked red in  Fig.~\ref{excitations}) are discussed in this work.  CEF excitations of Pr$^{3+}$ (marked green) were assigned based on the neutron scattering results for these two materials~\cite{Machida2005, Wen2017, Bonville2016} and are discussed in Ref.~[\onlinecite{Xu2021}]. Additionally, in the spectra of \pio\ we observe  electronic excitations at around 250~\cm\ (marked yellow) and excitations typical for iridates at around 650~\cm\ (marked blue)~\cite{Hasegawa2010}.

 \begin{table}[H]
 	\caption{Phonon modes assignment based on DFT calculation and experimental observation.}
 	\label{table:phonon}
 	\centering
 	\begin{ruledtabular}
 	\begin{tabular}[b]{>{\raggedright\arraybackslash}p{0.1\linewidth}									   >{\centering\arraybackslash}p{0.15\linewidth}
					   >{\centering\arraybackslash}p{0.18\linewidth}
					   >{\centering\arraybackslash}p{0.15\linewidth}
					   >{\centering\arraybackslash}p{0.18\linewidth}
					   }
		\multirow{2}{*}{Mode} & \multicolumn{3}{c}{ Frequency (\cm)} \\
			\cline{2-5}
		& \pio\ DFT & \pio\ Exp & \pzo\ DFT & \pzo\ Exp  \\
		\hline
		$T_{2g}^{(1)}$  & 306  & 297 (18.0) & 289 & 299 (6.7)\\
		\multirow{2}{*}{$E_g$ } & \multirow{2}{*}{318} & \multirow{2}{*}{334 (50.0)} & \multirow{2}{*}{305}  & 312 (9.8) \\
		& & & & 332 (17.1) \\
		$T_{2g}^{(2)}$ & 388 & 366 (17.1) & 391 & 350 (9.7) \\
		$A_{1g}$  & 487 & 502 (10.5) & 466 & 499 (7.2) \\
		$T_{2g}^{(3)}$ & 555 & - & 512 & 521 (9.4)  \\
		$T_{2g}^{(4)}$  & 633 & 612 (32.6) & 719 & - \\
	\end{tabular}
	\end{ruledtabular}
\label{frequencies}
 \end{table}

According to the symmetry analysis following the $Fd\bar{3}m$ symmetry of the unit cell, in the Raman spectra of \pzo\ and \pio\, there are six Raman-active phonons (see Table~\ref{table:wyckoff}). Note that Pr and Zr (Ir) are Raman inactive. Experimentally observed frequencies of the  phonons are in a good agreement with DFT calculations (see~Table~\ref{table:phonon}). In the Raman spectra of both materials, $A_{1g}$ and $T_{2g}$ phonon bands appear as sharp peaks, while $E_g$ phonons show more complicated spectra. The  phonon lines of \pio\ are  broadened  compared to \pzo\ spectra. For \pio, our experimental and calculated phonon spectra are in agreement with Ref.~[\onlinecite{Ueda2019}].

According to the calculations,  $E_g$ phonon frequency in \pzo\ spectra is expected at about 305~\cm. Instead,  we observe two bands at 312 and 332 \cm, which  are well resolved at low temperatures. Intensity of both bands  follows polarization dependence of an $E_g$ phonon in pyrochlores~\cite{Maczka2008,Ueda2019} (see also Fig.S1 in Supplemental Material (SM) \footnote{See Supplemental Material at [URL] for temperature dependence of line width and polarization dependence.}), but the phonon band at 332~\cm\  shows a weaker polarization dependence (see Fig.S1 in SM \cite{Note1}). The band of $E_g$ vibration in \pio\ spectra is found at about 334~\cm. It shows the largest width of all the $\Gamma$ phonons of about 55~\cm\ and very weak temperature dependence. While we cannot identify multiple components of this band as in \pzo\ spectra even at low temperatures, the large line width suggest an additional broadening or splitting into multiple partially overlapping components.

\begin{figure}[ht]
\centering
\includegraphics{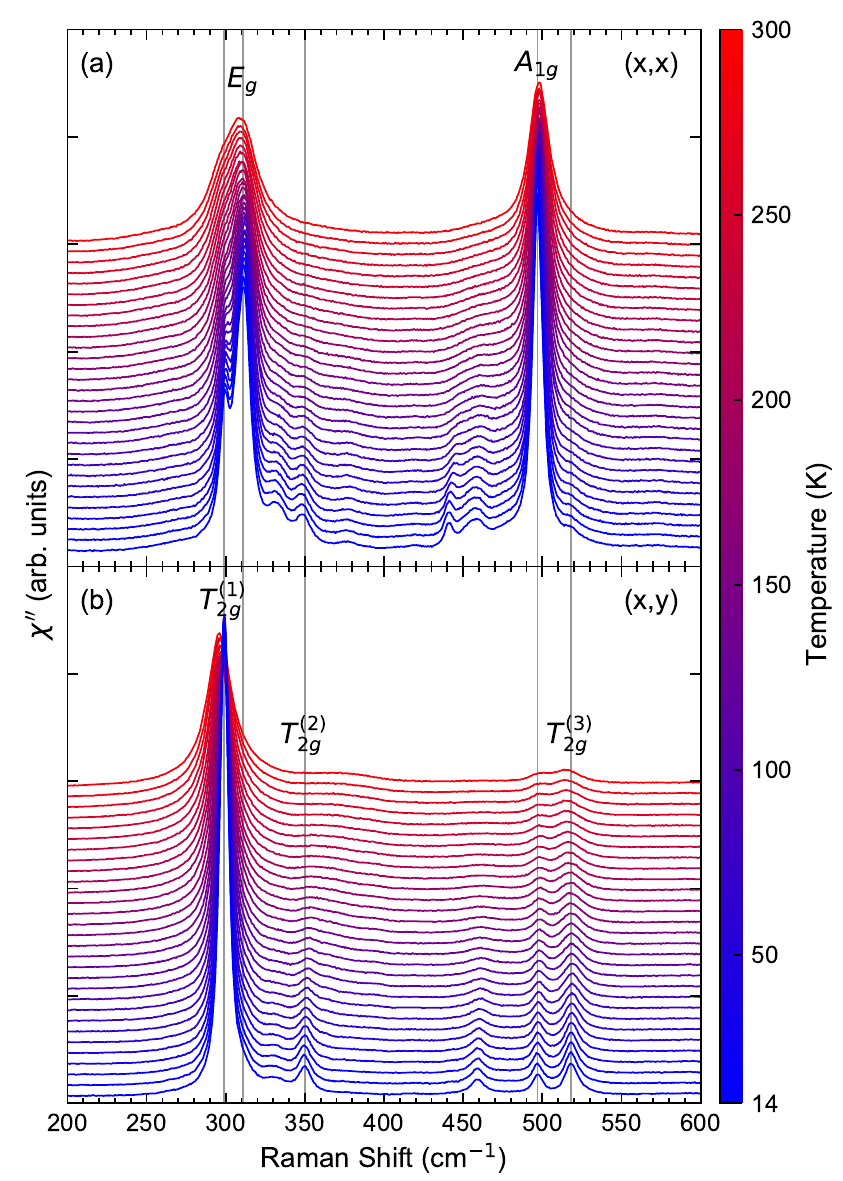}
\caption{Temperature dependence of the Raman scattering spectra of \pzo in (a) $(x,x)$ channel and (b) $(x,y)$ channel.}
\label{Fig2}
\end{figure}

We follow the temperature dependence of $A_{1g}$ and $T_{2g}$ phonons of \pzo\ and \pio. Fig.~\ref{Fig5} compares the temperature behavior of the widths and frequencies of the four lower frequency Raman-active phonons, which show the stronger temperature dependence. On a decrease of the sample temperature from 300~K, the phonons show hardening due to thermal contraction (see Fig.~\ref{Fig2}), except for $T^{(2)}_{2g}$ mode, which is discussed below. All phonons in \pzo\ show a conventional decrease of the line width due to phonon-phonon scattering \cite{Kim2012}, following the general formula
\begin{equation}
	\Gamma_{\mathrm{ph-ph}}(T) = \Gamma_0 + A(1 + 2n_B(\omega/2, T)),
\end{equation}
with $\Gamma_0$ of $5-7$ ~\cm, depending on the phonon.
The subtle effects of phonon-electron scattering in \pio\ become apparent when compared to the behavior of the \pzo\ phonons, see Fig.~\ref{Fig5}. At the lowest temperature all of the \pio\ phonons show larger width, while the temperature dependence of the width is less steep, and is described better by the scattering of phonons on interband transitions, as observed in semimetals~\cite{Osterhoudt2021},
\begin{equation}
	\Gamma_{\mathrm{ph-el}}(T) = \Gamma_0 + F(n_\mathrm{F}(\hbar\omega_a, T) - n_\mathrm{F}(\hbar\omega_a + \hbar\omega_{\rm{ph}}, T)).
\end{equation}

The  $T_{2g}^{(2)}$ phonon in the spectra of \pzo\ shows unconventional behavior different from the discussed above. In the spectral range of approximately $330-430$~\cm, the changes of the spectra on cooling reveal the $T_{2g}^{(2)}$ phonon as a sharp band only below 100~K (see Fig.~\ref{Fig4}a).  The frequency of 380~\cm\ corresponds to a CEF transition from the first excited $A_{1g}$ state at 77 \cm\ to the second excited state at about 440 \cm\ (see Fig.~\ref{Fig4}d). In the first approximation,  the total Raman intensity should be a superposition of CEF and phonon response, where CEF response will be changing as the population of the first excited $A_{1g}$ level at $E_1 = 77$~\cm, and the second excited at $E_2 = 440$ \cm. The spectral weight in this range can be calculated as $I(T) = \int^{330~\mathrm{cm}^{-1}}_{430~\mathrm{cm}^{-1}} \chi''(\omega) d\omega$ and is decreasing on cooling, indeed following a dependence of  $I(T)=A+B(e^{-E_1/k_\mathrm{B}T} - e^{-E_2/k_\mathrm{B}T})$ (see Fig.~\ref{Fig4}c). While the picture of a superposition of intensities works as a good approximation for the temperature dependence of the spectral weight, it does not explain the shape of the $T_{2g}^{(2)}$ phonon. In fact, the phonon appears as a weak antiresonance at the frequency of 371 \cm\ at room temperature, and gains the shape of a peak  as the spectral weight of the CEF decreases. The position of the $T_{2g}^{(2)}$ phonon defined as a position of an antiresonance or a maximum, depending on temperature, softens from 371 \cm\  down to 350~\cm\  on cooling from 300~K down to 14~K. The $A_{1g} \rightarrow E_g$ excitation is absent in \pio\ spectra, however, the $T_{2g}$ phonon  shows moderate softening from the frequency from  375~\cm\ at room temperature down to 364~\cm\ at 14~K.

\begin{figure}[ht]
\centering
\includegraphics{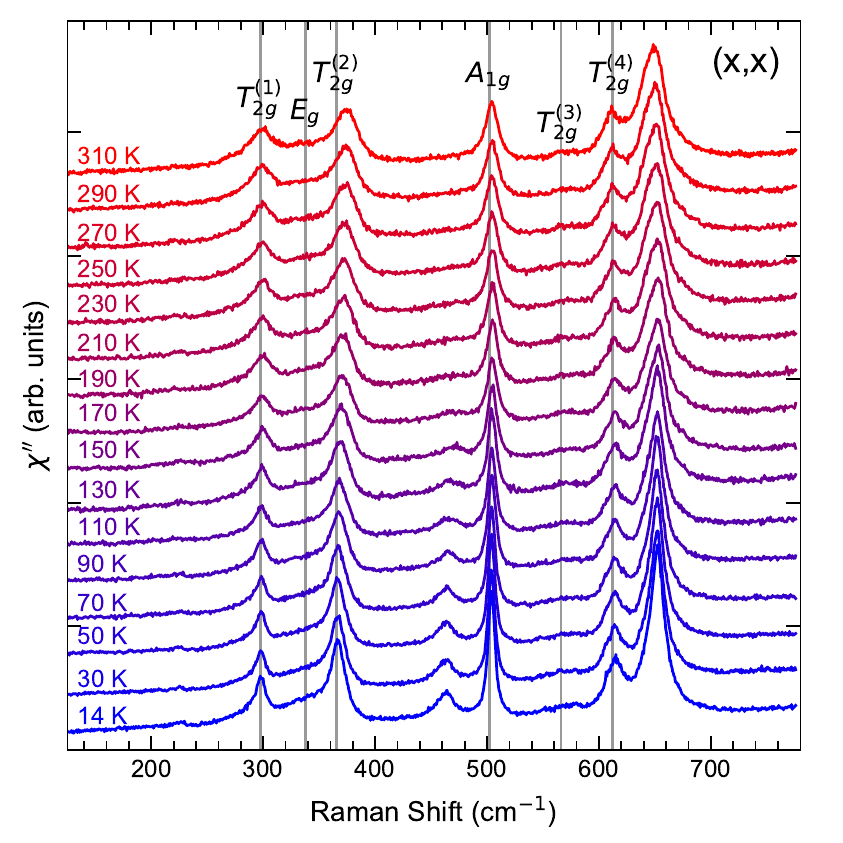}
\caption{Temperature dependence of the Raman scattering spectra of \pio.}
\label{Fig3}
\end{figure}

\begin{figure}[h]
\centering
\includegraphics[width=9cm]{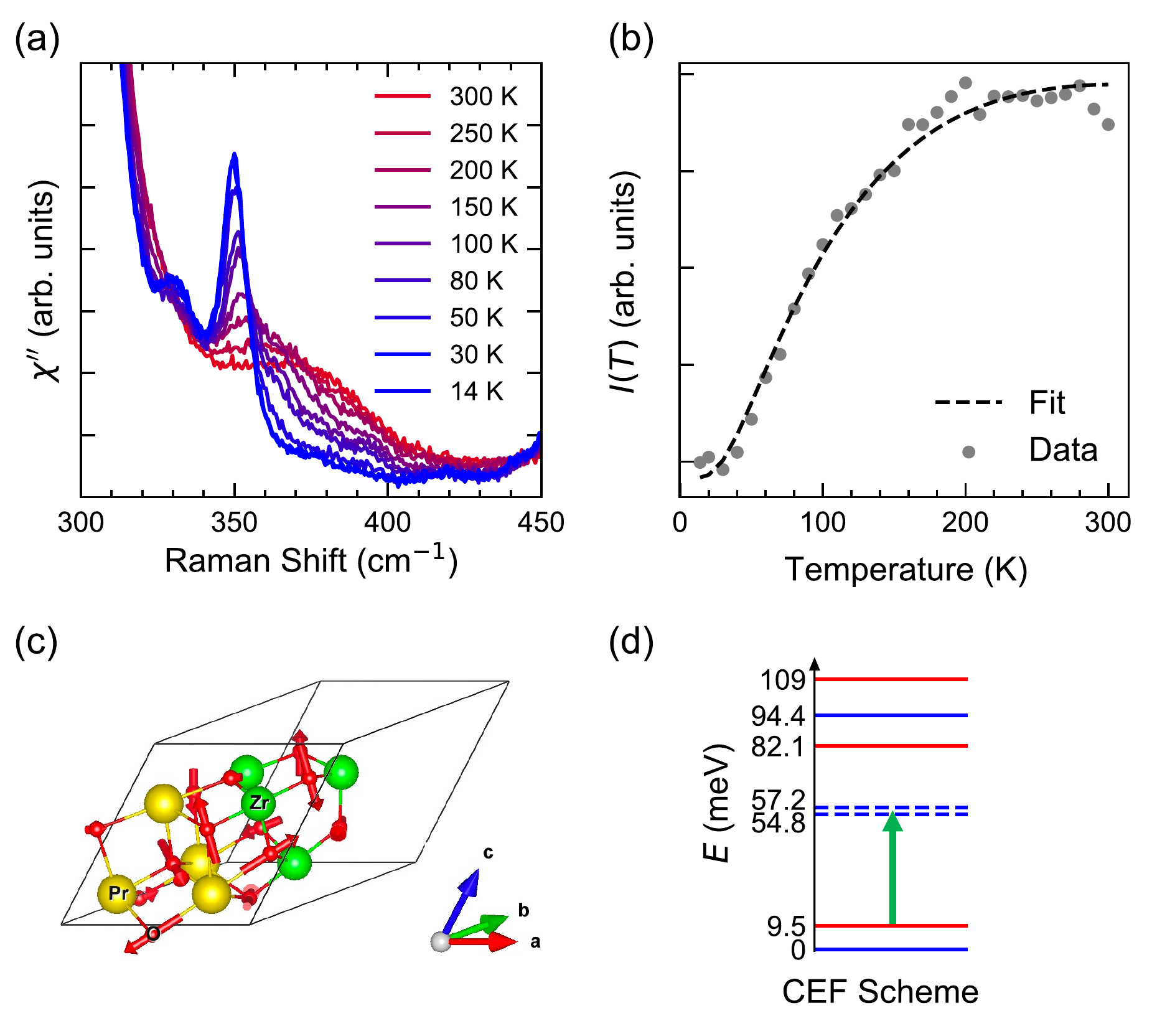}
\caption{(a) Temperature dependence of Raman scattering spectra of \pzo\ in the region of $A_{1g} \rightarrow E_g$ crystal electric field transitions in $(x,y)$ channel. (b) Temperature dependence of  spectral weight $I(T) = \int^{330~\mathrm{cm}^{-1}}_{430~\mathrm{cm}^{-1}} \chi''(\omega) d\omega$. (c) Atomic displacement of $T_{2g}^{(2)}$ phonon mode. (d) Crystal field scheme of Pr$^{3+}$ ions in \pzo\ crystal environment.}
\label{Fig4}
\end{figure}

\begin{figure}[ht]
\centering
\includegraphics[width=\linewidth]{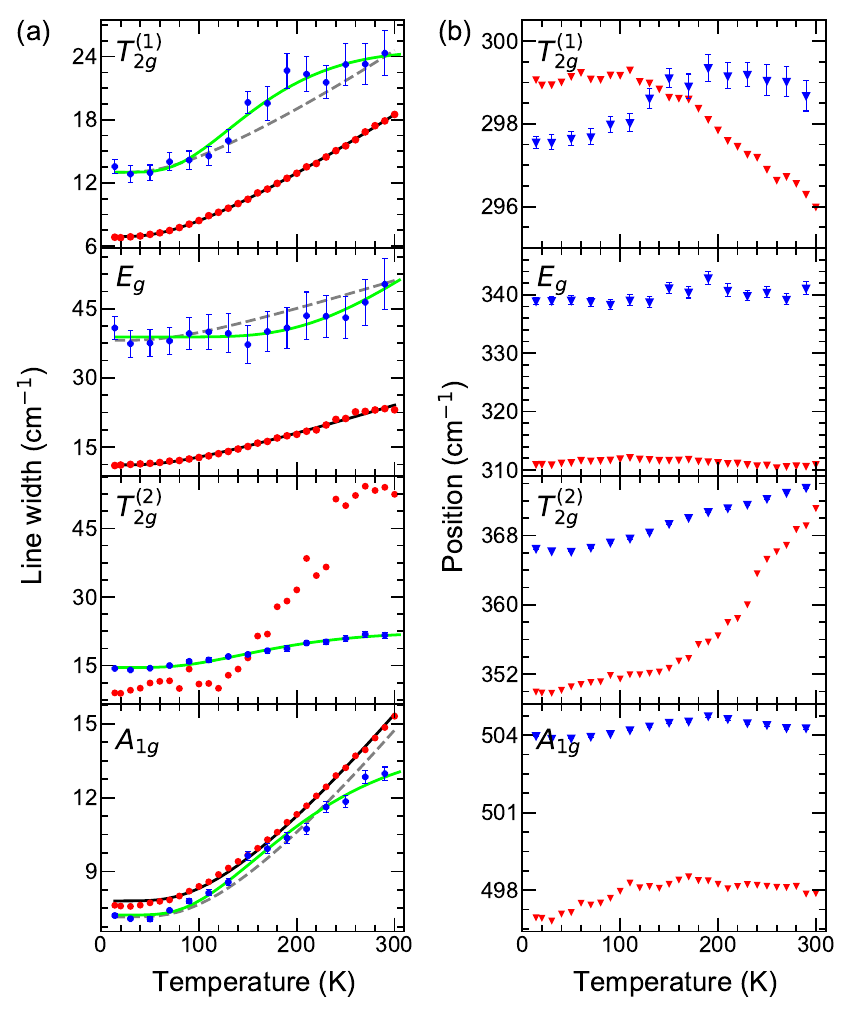}
\caption{Temperature dependence of (a) line widths and (b) positions for \pzo\ (blue) and \pio\ (red). Black lines are fits of line width of \pzo\ phonons with extended Klemens model. Dashed lines show the same temperature dependence with $\Gamma_0$ increased for \pio. Green lines are the fits of line width of \pio\ phonons by the phonon-electron scattering model. Note that the temperature dependence for \pio\ phonons is best fit by phonon-electron scattering. Green lines were fit by the phonon-electron scattering model for \pio. }
\label{Fig5}
\end{figure}

 \section{Discussion}

\subsection{Phonon-electron scattering}

The energies of the phonon modes are very similar between \pio\ and \pzo, as expected for materials with similar crystal structures, which differ only by a substitution of $B$ position in the $A_2B_2$O$_7$ of pyrochlore structure. However, the difference both in low-temperature width of the phonons, and in their temperature dependence is considerable. At low temperatures, all the phonon bands of \pio\ show larger line width than that of \pzo. We assign this difference not to the larger random disorder in \pio, but to the effects of phonon-eletron coupling. It was shown \cite{Coulter2019,Osterhoudt2021} that in semimetals, phonon can scatter on the direct ($q = 0$) interband transitions, producing a temperature dependence of width very different from the Klemens' model. This can explain the temperature behavior of the phonons in \pio\ ~(Fig.~\ref{Fig5}). The fact that two phonons of different symmetries, $T_{2g}$ and $A_{1g}$, which are not coupled to magnetic degrees of freedom, show the evidence of phonon-electron scattering, suggests that the scattering occurs on the interband transitions away from the highest symmetry points of the BZ, and is not restricted by symmetry. The high probability of such scattering could be explained based on the calculated \pio\ band structure, which suggests a band touching at the $\Gamma$ point of BZ, and a narrow gap between nearly parallel bands between $\Gamma$ and $L$ points~\cite{Kondo2015, Ishii2015}, while the calculations Ref.~[\onlinecite{Zhang2017}] suggest these parallel bands below the Fermi energy. The gap in the range of $300-500$~\cm\ ($37-65$~meV) between these bands would provide a suitable phase space for phonon-electron scattering on interband transitions. Moreover, the presence of this scattering testifies for the existence of such a band gap. An interesting  proposal of an increase of conductivity via electron-phonon scattering on acoustic modes~\cite{Osterhoudt2021} could be also relevant to \pio.

\subsection{Control of the super-exchange values through structure}
The information about the phonons and their temperature dependence supply us the knowledge of the subtle effects of  magneto-elastic interactions, and the importance of the structure for magnetic behavior.
Here we will focus on the behavior of the $T^{(2)}_{2g}$ phonon found at about 350~\cm\, which involves the change of Pr-O-Pr angle, which  determines the size of the  super-exchange interactions between Pr atoms. In the spectra of both materials, the $T^{(2)}_{2g}$ phonon is softening on cooling (see Fig.~\ref{Fig5}). However, the  DFT calculations do not find any significant magneto-elastic coupling for this phonon.

We suggest that the phonon behavior reflects the changes of the lattice on thermal contraction. This effect is much larger in \pzo\ (20 \cm\ shift) compared to \pio\ (8 \cm\ shift), where magnetic interactions are more complex due to the magnetism of Ir. Softening of the phonon can evidence the change of Pr-O-Pr angle that in turn leads to the decrease of the super-exchange values. Indeed, a decrease of super-exchange below 100~K in \pzo\ was suggested as an interpretation of the temperature dependence of  magnetic susceptibility~\cite{Bonville2016}. A similar but weaker tendency is observed for \pio\ \cite{Nakatsuji2006}.

In the spectra of \pzo, the behavior of the $T_{2g	}$ phonon is more involved than a pure softening. At temperatures above about 100~K, this phonon shows very strong changes associated with the decrease of CEF $A_{1g} \rightarrow E_g$ intensity due to the depopulation of the $A_{1g}$ level. Moreover, at room temperature the phonon appears as an antiresonance in the broad band of CEF $A_{1g} \rightarrow E_g$ transition, which is  an evidence of an interaction of the phonon with this transition~\cite{Fano1961}. The observed spectroscopic effect is very different from vibronic effects on mixing of CEF and phonons, discussed in ~\cite{Thalmeier1984, Xu2021}.

\subsection{$E_g$ phonons}

Anomalous behavior was also detected from $E_g$ phonons, which show a double band in the spectra of \pzo, and an increased width in the spectra of \pio.

$E_g$ degree of freedom is sensitive to magneto-elastic coupling in pyrochlore materials as revealed on magnetic phase transitions \cite{Ueda2019}. However, both compounds are known not to order magnetically, and the properties of the $E_g$ phonons do not show any pronounced temperature dependence. A possible interpretation of a splitting of a doubly degenerate band observed in the whole temperature range is a local symmetry breaking. For the highest quality crystals of \pzo\ free from oxygen vacancies and site mixing \cite{Tang2020}, a shift of Pr ion from a central position~\cite{Martin2017,Koohpayeh2014}, and a local tilting of octahedra~\cite{Trump2018}, which reduces local symmetry from $Fd\bar3m$ (point group $O_h$) to $P4_32_12$ (point group $D_4$) is discussed. With this symmetry reduction, $E_g$ phonons split into two in-plane modes of $B_1$ symmetry (see Table.~\ref{table:correlation}).  $T_{2g}$ phonons  split into $B_2$ and $E$ modes, where the $E$ modes will be observed in $xz$ and $yz$ channels which can be only detected in  out-of-plane scattering, and thus will not appear in the spectra in $(x,y)$  and $(x^2-y^2)$ channels. This symmetry considerations can explain why $T_{2g}$ phonons do not show any anomalous behavior of their width.

The large width of $E_g$ phonons in \pio\ spectra, which appear to be about twice the width of the other phonons, suggest that a splitting of this band is also present in \pio, but cannot be resolved due to the large band width, and can have a similar origin.

 \begin{table}
 	\centering
 	\caption{Correlations between $O_h$ and $D_{4}$ groups and symmetries.}
 	\label{table:correlation}
 	\begin{ruledtabular}
 	\begin{tabular}{
 		>{\centering\arraybackslash}p{0.05\linewidth}
 		>{\centering\arraybackslash}p{0.3\linewidth}
 		>{\centering\arraybackslash}p{0.05\linewidth}
 		>{\centering\arraybackslash}p{0.2\linewidth}
 		>{\centering\arraybackslash}p{0.3\linewidth}
 		}
		\multicolumn{2}{c}{$O_h$} & & & $D_{4}$ \\
		\hline
		$A_{1g}$ & ($\bm{x^2{+}y^2}{+}z^2$) & & & \tikzmark{d}$A_{1}~(\bm{x^2{+}y^2}$, $z^2$)\\
                                      &  & &  \\
		\multirow{2}{*}{$E_g$}\tikzmark{a} & ($\bm{x^2{-}y^2}$ &  & \tikzmark{b}\multirow{2}{*}{$A\tikzmark{c}_{1}{+}B\tikzmark{e}_{1}$} &  $\tikzmark{f}B_{1}~(\bm{x^2{-}y^2})$\\
		& $2z^2{-}\bm{x^2-y^2})$ & & 	&     \\
                                     &  & &   & $B_{2}~(\bm{xy})$ \\
		$T_{2g}$ & $(\bm{xy},xz,yz)$ & & $E{+}B_{2}$ & $E~(xz, yz)$
 	\end{tabular}
 	\end{ruledtabular}
 \end{table}

 \begin{tikzpicture}[overlay, remember picture, yshift=.25\baselineskip, shorten >=.5pt, shorten <=.5pt]
    \draw [->] ([xshift=70pt, yshift=-3pt]{pic cs:a}) -- ([yshift=-3pt]{pic cs:b});
    \draw [->] ([yshift=5pt]{pic cs:c}) -- ({pic cs:d});
    \draw [->] ([yshift=4pt]{pic cs:e}) -- ([yshift=3pt]{pic cs:f});
	\draw [->] ([xshift=70pt, yshift=-29pt]{pic cs:a}) -- ([yshift=-29pt]{pic cs:b});
  \end{tikzpicture}

\section{Conclusions}

In this work we compare the Raman phonon spectra of \pzo\ and \pio, which have a very similar crystal structure, but drastically different electronic properties. Raman phonons observed at similar frequencies demonstrate  different behavior with temperature, determined by phonon-phonon scattering in the case of the insulating \pzo, and phonon-electron scattering in the case of \pio. A softening of the phonon changing Pr-O-Pr angle on cooling explains the decrease of magnetic interactions $J$ deduced previously from magnetic susceptibility measurements. The broadening of the $E_g$  phonons demonstrates the presence of the  disordered tilting of tetrahedra suggested for pyrochlore materials.  As a whole, our study elucidates many subtle questions about electronic and magnetic properties of these materials through the behavior of the underlying lattice.

\section{Acknowledgements}
The authors are thankful to C.~Broholm, S.~Bhattacharjee, T.~McQueen, J-J.~Wen, O.~Tchernyshev, and H.~Zhang for useful discussions. This work was supported as part of the Institute for Quantum Matter, an Energy Frontier Research Center funded by the U.S. Department of Energy, Office of Science, Basic Energy Sciences under Award No.~DE-SC0019331. This work in Japan is partially supported by CREST (Grant Number: JPMJCR18T3 and JPMJCR15Q5), by New Energy and Industrial Technology Development Organization (NEDO), by Grants-in-Aids for Scientific Research on Innovative Areas (Grant Number: 15H05882 and 15H05883) from the Ministry of Education, Culture, Sports, Science, and Technology of Japan, and by Grants-in-Aid for Scientific Research (Grant Number: 19H00650).
\bibliography{./PZO_PIO}

\end{document}